\documentclass[twoside,english,a4,doublecol]{epl2}

\usepackage{bm}
\usepackage{graphicx}
\usepackage{amsmath}
\usepackage{amssymb}

\usepackage[normalem]{ulem}
\newcommand{\tb}{\mathring{\tau}}
\newcommand{\cb}{\mathring{c}}
\newcommand{\rmd}{\mathrm{d}}
\newcommand{\rme}{\mathrm{e}}
\newcommand{\rmi}{\mathrm{i}}
\newcommand{\pint}{\int\frac{\rmd^{d-1}p}{(2\pi)^{d-1}}}
\newcommand{\fex}{f_{\mathrm{ex}}}
\newcommand{\Leff}{L_{\mathrm{eff}}}
\newcommand{\fb}{f_{\mathrm{b}}}
\newcommand{\rb}{r_{\mathrm{b}}}
\newcommand{\fs}{f_{\mathrm{s}}}
\newcommand{\mat}[1]{\mathbf{#1}}
\makeatletter
\newcommand{\Tr}{\mathop{\operator@font tr}\nolimits}
\newcommand{\arsinh}{\mathop{\operator@font arsinh}\nolimits}
\makeatother

\newcommand{\aseq}{\simeq}
\newcommand{\asprop}{\sim}
\newcommand{\etwa}{\approx}

\begin{document}


\title{Exact thermodynamic Casimir forces for an interacting three-dimensional model~system in film geometry with free surfaces} 
\shorttitle{Exact thermodynamic Casimir forces}

\author{H.~W. Diehl\inst{1} \and Daniel Gr\"uneberg\inst{1} \and Martin Hasenbusch\inst{2} \and Alfred Hucht\inst{1} \and Sergei B. Rutkevich\inst{3,1} \and Felix M. Schmidt\inst{1} }

\shortauthor{H.~W. Diehl \etal}

\institute{
\inst{1} Fakult\"at f\"ur Physik, Universit\"at Duisburg-Essen, D-47058 Duisburg, Germany\\
\inst{2} Institut f\"ur Physik, Humboldt-Universit\"at zu Berlin, Newtonstr.~15, D-12489 Berlin, Germany\\
\inst{3} Institute of Solid State and Semiconductor Physics, Minsk, Belarus
}


\abstract{
The limit $n\to\infty$ of the classical $O(n)$ $\phi^4$ model on a 3d film with free surfaces is studied. Its exact solution involves a selfconsistent 1d Schr\"odinger equation, which is solved numerically for a partially discretized as well as for a fully discrete lattice model. Extremely precise results are obtained for the scaled Casimir force at all temperatures. Obtained via a single framework, they exhibit all relevant qualitative features of the thermodynamic Casimir force known from wetting experiments on $^4$He and Monte Carlo simulations, including a pronounced minimum below the bulk critical point.
}

\maketitle

A celebrated example of fluctuation-induced forces is the Casimir force between two metallic, grounded plates in vacuum \cite{Cas48}.\footnote{For a review of the {C}asimir effect in {QED} and an extensive
  lists of references, see \cite{casiqmrev}} Such forces caused by the confinement of quantum electrodynamics (QED) vacuum fluctuations of the electromagnetic fields are expected to have considerable technological relevance. This has made them the focus of much ongoing research activity. During the past two decades,  it has become increasingly clear that a wealth of similarly interesting classical analogs of such effective forces, induced by thermal rather than quantum fluctuations, exist \cite{FdG78}.\footnote{For reviews of the thermodynamic Casimir effect and extensive lists of references, see \cite{casitdrev}}
Two  important classes of such ``thermodynamic Casimir forces''\footnote{Following established conventions we use the term "thermodynamic Casimir forces" for forces induced by thermal fluctuations, in particular, also for near-critical Casimir forces, reserving the name critical Casimir forces to those where the medium is at a critical point. This topic must not be confused with those of thermal effects on QED Casimir forces and thermal Casimir-Polder forces, which are less universal since material properties of  the media and confining objects matter; see, e.g., \cite{tdc}} are forces induced by fluctuations  in  nearly (multi)critical media between immersed macroscopic bodies or boundaries, and forces due to confined Goldstone modes \cite{KG99}. 
Clear experimental evidence for the existence of such thermodynamic Casimir forces was provided first indirectly by measurements of the thinning of ${}^4$He wetting films at the $\lambda$-point as the temperature $T$ is lowered below the bulk critical temperature $T_\mathrm{c}$ \cite{GC99}. Subsequently, direct measurements of the thermodynamic Casimir force on colloidal particles in binary liquids near the consolute point could be achieved \cite{HHGDB08}.

Despite obvious analogies, crucial qualitative differences between thermodynamic and QED Casimir forces exist. First, the latter usually can be studied in terms of effective \emph{free} field theories in confined geometries where the interaction of the electromagnetic field with the material boundaries is taken into account through boundary conditions. By contrast, investigations of
thermodynamic Casimir forces at (multi)critical points necessarily involve \emph{interacting} field theories. Second, whereas electromagnetic fields average to zero in the ground state, the thermal averages $\langle\bm{\phi}(\bm{x})\rangle$ of fluctuating densities $\bm{\phi}(\bm{x})$ (order parameters) associated with thermodynamic  Casimir forces do not necessarily vanish. Nonzero profiles $\langle\bm{\phi}(\bm{x})\rangle$ can occur at all temperatures $T$  when the symmetry $\bm{\phi}\to -\bm{\phi}$ is explicitly broken (as it generically is for fluids and binary fluid mixtures in contact with walls), and  at low temperatures if this symmetry is spontaneously broken. When a nonzero profile exists, it will respond to a change of the  separation between boundaries and hence cause an effective force even in the absence of fluctuations. The presence of such nonfluctuating background contributions to thermodynamic Casimir forces substantially spoils the analogy with the QED case. Obviously, thermodynamic Casimir forces in $^4$He films provide much better analogs than those in binary mixed fluids because neither a spontaneous breakdown of the $U(1)$ symmetry is possible for finite thickness $L$ of the film, nor an explicit breakdown of the symmetry through the boundary planes.

It is therefore very unfortunate that no viable theory other than sophisticated Monte Carlo simulations of lattice XY models \cite{MC1,MC2,MC3} has emerged which is capable to yield, within a single framework, all experimentally observed \cite{GC99} relevant qualitative features of the reduced Casimir force $\beta \mathcal{F}_\mathrm{C}(T,L)$, where $\beta=1/k_\mathrm{B}T$. For a $d$-dimensional film, the latter takes the scaling form $\beta \mathcal{F}_\mathrm{C}(T,L) \aseq L^{-d}\,\vartheta(x)$ 
with scaling variable $x=t(L/\xi_+)^{1/\nu}$ on large length scales\footnote{Throughout this work, the symbol $\aseq$ means "asymptotically equal" in the respective limit, e.g., $f(L)\aseq g(L)\Leftrightarrow\lim_{L\rightarrow\infty}f(L)/g(L)=1$.}, where $t=T/T_\mathrm{c}-1$ while $\nu$ and $\xi_+$ are the critical exponent and $t>0$ amplitude of the bulk correlation length, respectively. The observed features of the scaling function $\vartheta(x)$ are: (i) $\vartheta(x)<0$ for all $x$, (ii) a relatively small critical value $\vartheta(0)$, (iii) a smooth minimum at $x_{\mathrm{min}}<0$, and (iv) a nonvanishing $T\to 0$ limit $\vartheta(-\infty)$. A theory based on renormalization-group (RG) improved Landau theory \cite{ZSRKC07,MGD07}, though giving some insight, suffers from severe deficiencies: it erroneously predicts (d1) an ordered low-temperature phase for $d=3$ and $L<\infty$,  (d2) a much to deep minimum of $\vartheta$ whose derivative has a (d3) jump discontinuity there, and (d4) a vanishing $T\to 0$ limit $\vartheta(-\infty)$.

\begin{figure}[t]
\includegraphics[width=0.95\columnwidth]{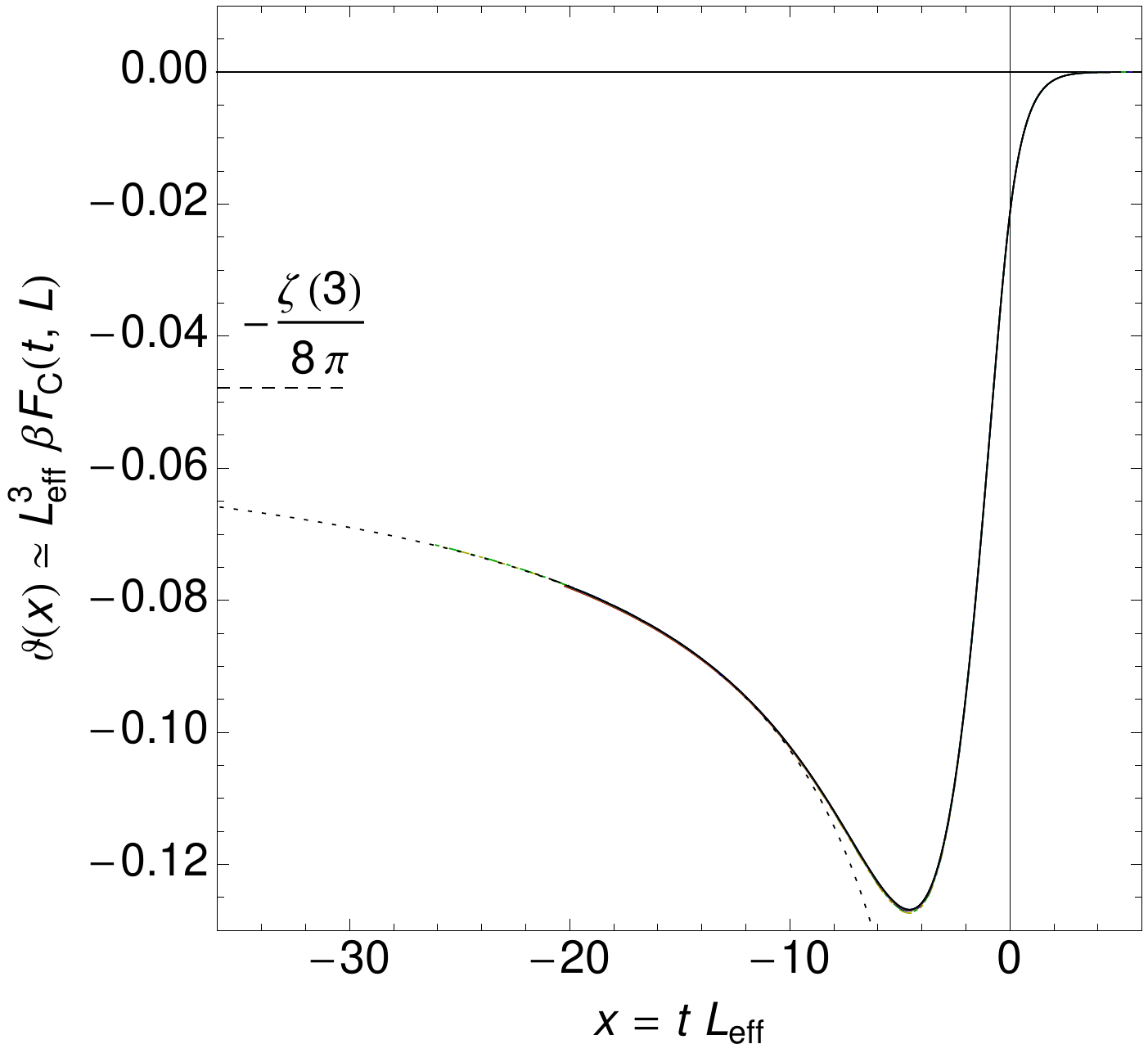}
\caption{(Color online) Scaling function $\vartheta(x)$ of the critical Casimir force for $L = \{ 65, 97, 129, 193, 257 \}$ (model A, solid lines) and $L = \{ 97, 129, 193, 257 \}$ (model B, dash-dotted lines, for model definition see text). The dashed horizontal line indicates the Goldstone value $\vartheta(-\infty)=-\zeta(3)/8\pi$, and the dotted curve represents the limiting behavior
$\vartheta(x)\aseq\vartheta(-\infty)+ (c \ln|x| + d)/x$, with $c=0.07856(5)$ and $d=0.3673(5)$. 
The effective thickness $\Leff$ accounts for leading scaling corrections (see text).
Note that deviations from the excellent data collapse are only visible under magnification.
}
\label{fig:theta}
\end{figure}
In this Letter, we report exact results for the Casimir force of the $O(n)$ $\phi^4$ model on a  film $\mathbb{R}^2\times [0,L]$ with free surfaces at $z=0$ and $z=L$ in the limit $n\to\infty$. Its $n=2$ analog describes $^4$He fluid films near the $\lambda$-point. The model has, for general $n\geq2$, an ordered low-$T$ bulk ($L=\infty$) phase. For finite $L$, long-range order is restricted to $T=0$ since low-energy (spin wave) excitations destroy long-range order at any temperature $T>0$. Further, the presence of confined Goldstone modes at $T=0$ implies a nonzero $T\to 0$ limit of the Casimir force. Owing to the breakdown of translation invariance along the $z$-direction perpendicular to the surfaces, the exact $n\to\infty$ solution does not correspond to a  mean spherical model with global constraint \cite{Kno73}, but involves a 1d Schr\"odinger equation with a selfconsistent potential $V(z)$. The exact scaling function $\vartheta(x)$ for $n=\infty$ can be expressed in terms of the eigenvalues and eigenfunctions of this equation. Determining them by numerical means, we managed to get the extremely precise results displayed in Fig.~\ref{fig:theta}. Determined within a single theoretical framework, these exhibit all features (i)--(iv) mentioned above but none of the deficiencies (d1)--(d4).

Our best estimate for the Casimir amplitude  is
\begin{equation}\label{eq:Deltaest}
\Delta_{\mathrm{C}}=\vartheta(0)/2=-0.01077340685024782(1).
\end{equation}
The  function $\vartheta(x)$ has a pronounced minimum 
$\vartheta_\mathrm{min}=-0.1268565841360(1)$ at $x_\mathrm{min}=-4.55702477008(1)$.
In the low-temperature limit $x\to-\infty$, it approaches the value $-\zeta(3)/8\pi=-0.0478283245\ldots$ \cite{freeNbc} of a massless free theory with Neumann boundary conditions.\footnote{Arguments similar to those used in \cite{lowTNbc} show that the low-$T$ limit is described by a nonlinear $\sigma$ model on a film with Neumann boundary conditions.}
Using an appropriate low-$T$ model --- a nonlinear $\sigma$ model on a film with Neumann boundary conditions \cite{lowTNbc} --- we can show that the asymptotic  behavior is of the form
$\vartheta(x)-\vartheta(-\infty) \aseq c\,x^{-1}\ln |x|$ \cite{we}.

Both the form of these logarithmic anomalies and  the existence of a minimum at $x_{\mathrm{min}}<0$ are intimately related to the breaking of translational invariance across the slab. To appreciate this, one should note the following. As may be gleaned from \cite{BM77} and will be shown below, the exact $n\to\infty$ solution reduces to a constrained Gaussian model with an effective $\phi^2/2$ term whose interaction constant $\tb-V(z)$ involves a selfconsistent potential $V(z)$. According to \cite{BM77}, the $t=0$ analog of $V(z)$ for the semi-infinite case $L=\infty$ is given by $-1/4z^2$ \cite{BM77}. This algebraic behavior carries over to the near-boundary behavior of $V(z)$ for $t<0$ on scales small compared to both the Josephson coherence length $\propto |t|^{-\nu}$ and $L$ \cite{we}. By contrast, the counterpart of $V(z)$ for periodic boundary conditions is independent of $z$, an enormous simplification which enables one to determine the corresponding scaling function $\vartheta^{\mathrm{pbc}}(x)$ in closed analytical form \cite{Dan98}. It decreases quickly and monotonically from zero to its Goldstone value $\vartheta^{\mathrm{pbc}}(-\infty)=-\zeta(3)/\pi$, has no minimum at finite $x<0$, and approaches the $x\to-\infty$ limit $\asprop |x|\,\rme^{-|x|}$.

We next turn to an outline of the essentials of our calculations.
The model is described by the Hamiltonian
\begin{eqnarray}\label{eq:Ham}
\mathcal{H}&\!\!\!\!\!=\!\!\!\!\!&\int\rmd^{d-1}y\bigg\{\int_0^L\rmd{z}\;\Big[\frac{1}{2}(\nabla\bm{\phi})^2+\frac{\mathring\tau}{2}\phi^2+\frac{g}{4! n}\phi^4\Big]\nonumber\\
&&\strut+\frac{\mathring{c}_1}{2}\phi^2(\bm{y},0)+\frac{\mathring {c}_2}{2}\phi^2(\bm{y},L)\bigg\}
\end{eqnarray}
with $d=3$, where $\bm{\phi}=(\phi_\alpha)$ is an $n$-component field and $\bm{y}\in\mathbb{R}^{d-1}$ denotes the lateral coordinates. The boundary terms entail the boundary conditions 
$(\partial_z- \mathring{c}_1)\bm{\phi}|_{z=0}$ and $(\partial_z+\mathring{c}_2)\bm{\phi}|_{z=L}$. 

The $n\to\infty$ limit can be derived by standard means. Upon making a Hubbard-Stratonovich transformation,  the partition function $Z=\int\mathcal{D}[\bm{\phi}]\,\rme^{-\mathcal{H}}$ can be written as
\begin{equation}
Z \propto \! \int \! \mathcal{D}[\bm{\phi}] \mathcal{D}[\psi]\,\rme^{-\frac{1}{2}\int\rmd^{d-1}y\int_0^L\rmd{z}[\bm{\phi}(\tb+\rmi\psi-\nabla^2)\bm{\phi}-\frac{3n}{g}\psi^2]},
\end{equation}
where the Laplacian $\nabla^2$ is subject to the above-mentioned boundary conditions. In the limit $n\to\infty$, the $\psi$-integral can be evaluated by saddle-point integration. Writing the saddle point as $\rmi\psi_0(z)=V(z)-\tb$,  the reduced free energy per area $A=\int\rmd^{d-1}y$, $f_L\equiv -(An)^{-1}\ln Z$, becomes
\begin{eqnarray}\label{eq:f}
f_L&\!\!\!\!\!=\!\!\!\!\!&\frac{1}{2}\int_{0}^{L}\rmd{z}\bigg\{\pint\langle z|\ln(\bm{p}^{2}-\partial_{z}^{2}+V)|z\rangle\nonumber \\ &&\strut -\frac{3}{g}{[\tb-V(z)]}^{2}\bigg\}+f_L^{(0)}\,,
\end{eqnarray}
where $f_L^{(0)}$ is a trivial background term which does not matter henceforth.

The stationarity condition $\delta f_L/\delta V(z)=0$ yields 
\begin{equation}\label{eq:V-tau}
\tb-V(z)=-\frac{g}{6}\pint\sum_{\nu}\frac{|\varphi_{\nu}(z)|^2}{\bm{p}^{2}+\epsilon_{\nu}},
\end{equation}
where $\epsilon_\nu$ and $\varphi_\nu(z)=\langle z|\nu\rangle$ are the eigensolutions of 
\begin{equation}\label{eq:ev}
[-\partial_{z}^{2}+V(z)]\varphi_{\nu}(z)=\epsilon_{\nu}\varphi_{\nu}(z).
\end{equation}

The $\bm{p}$-integrals  on the right-hand sides of Eqs.~\eqref{eq:f} and \eqref{eq:V-tau} are ultraviolet (UV) divergent at $d=3$. To make the model (\ref{eq:Ham}) well defined 
and suitable for numerical calculations,  we must regularize these divergences.
We study two distinct regularized versions of model \eqref{eq:Ham}.
In the first (A), 	only the $z$-coordinate is discretized, and  the $\bm{p}$-integrals are regularized dimensionally. In the second (B), a fully discrete lattice model is investigated. 

{\it Model A ---}
The system consists of $L$ layers located at $z=1,\cdots,L$, where we 
replace the operator $\partial_z^2$ in Eq.~\eqref{eq:ev} by its  discrete analog, the $L{\times}L$ matrix $\mat{D}^2=(-2\delta_{z,z'}+\delta_{|z-z'|,1})$.  Rather than including analogs of the boundary terms $\propto \cb_j$, we impose the Dirichlet boundary conditions $\bm{\phi}|_{z=0}=\bm{\phi}|_{z=L+1}=\bm{0}$. 
The thickness change $L\to L+1$ is accounted for in the numerical analysis by introducing an effective thickness $\Leff$ (see below).
The Hamiltonian of Eq.~\eqref{eq:ev} becomes the matrix $\mat{H}=-\mat{D}^2+\mat{V}$ with 
$\mat{V}=\mathrm{diag}(V_z)$.\footnote{A preliminary analysis of the critical case was made in \cite{Cometal}} The dimensionally regularized $\bm{p}$-integral in Eq.~\eqref{eq:V-tau} is straightforward. It produces a simple pole at $d=3$, which gets absorbed in the bulk critical value $\tb_c$. To see this, we subtract from Eq.~\eqref{eq:V-tau} its bulk analog, making the appropriate replacements $\sum_\nu\ldots|\varphi_{\nu z}|^2\to \int_{-\pi}^\pi\frac{\rmd{k}}{2\pi}$ and $\epsilon_\nu\to \epsilon(k)=\rb+4\sin^2(k/2)$, where $\rb$ is the inverse bulk susceptibility. Writing $\tb=\tb_c+\tau$ and noting that $\rb=0$ at $\tb_c$, we can set $d=3$ to obtain
\begin{subequations}\label{eq:sctfex}
\begin{equation} \label{eq:tauVj}
24 \pi g^{-1} (\tau-V_z) = \langle z |\ln\mat{H}| z\rangle.
\end{equation}

To eliminate the UV singularities of the bulk free energy density $\fb=\lim_{L\to\infty}f_L/L$ we subtract from it its Taylor expansion to first order in $\tau$. This gives an UV finite renormalized $\fb(\tau,g)$ which at $d=3$ becomes
\begin{eqnarray}\label{eq:fb}
\fb(\tau,g)&\!\!\!\!\!=\!\!\!\!\!&\frac{1}{8\pi} \sqrt{\rb(4+\rb)}-\frac{2+\rb}{4\pi}\arsinh\sqrt{\rb/4}\nonumber\\ &&\strut -\frac{3}{2g}(\tau-\rb)^2,
\end{eqnarray}
where $\rb$ is the solution to $\rb=\tau-\frac{g}{12\pi}\arsinh\sqrt{\rb/4}$ or $0$ depending on whether $\tau\ge0$ or $<0$.
The chosen bulk counterterms also absorb  the UV singularities of $f_L$. Their contributions cancel in the excess free energy $\fex=f_L-L\fb$, yielding the UV finite $d=3$ result \cite{we}
\begin{eqnarray}\label{eq:fex}
\fex(\tau,g,L)&\!\!\!\!\!=\!\!\!\!\!&\frac{1}{8\pi}\Tr[\mat{H}(1 - \ln\mat{H})]-\frac{3}{2g}\Tr[(\tau-\mat{V})^2]\nonumber\\ &&\strut-L\fb(\tau,g).
\end{eqnarray}
\end{subequations}

\begin{figure}[t]
\includegraphics[width=0.9\columnwidth]{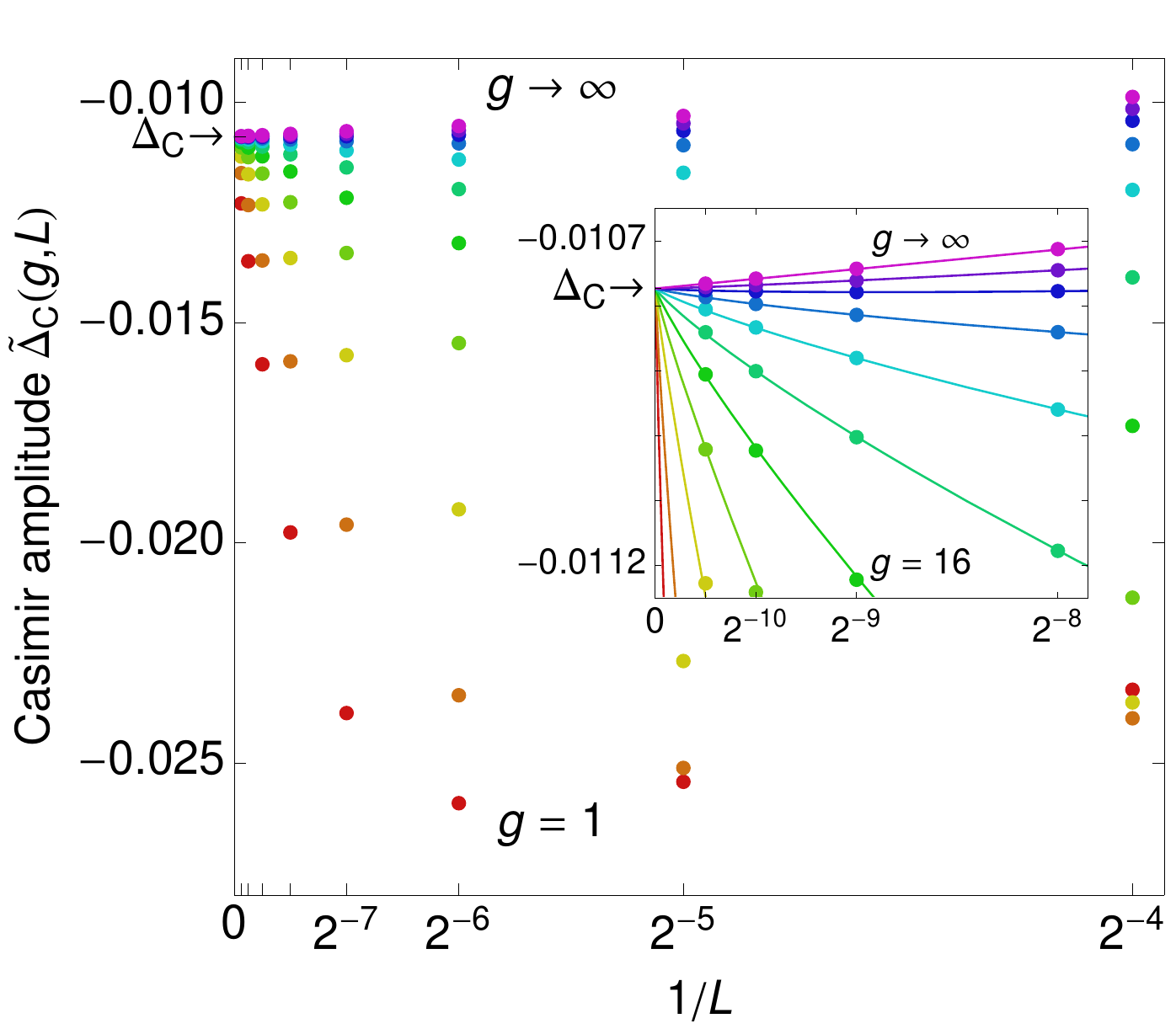}
\caption{(Color online) Effective Casimir amplitude $\tilde{\Delta}_\mathrm{C}(g,L)$ for different values of $g = \{ 1,2,4,\ldots,512,\infty \}$. The solid lines are fits including corrections  $\propto L^{-1} \ln L$, which vanish as $1/g$ for $g\to\infty$ (see text). The dotted lines are guides to the eyes.}
\label{fig:deltag}
\end{figure}

Solving Eq.~\eqref{eq:tauVj} numerically first for $\tau=0$, we compute $\fex(0,g,L)$ to determine the surface free energy $\fs(0,g)=\fex(0,g,\infty)/2$ and  the Casimir amplitude $\Delta_{\mathrm{C}}\equiv \lim_{L\to\infty}L^2[\fex(0,g,L)-2\fs(0,g)]$. To extract precise values from the data, knowledge about corrections to scaling is important. Clearly, the usual Wegner corrections governed by the $n\to\infty$ exponent $\omega=4-d$ must be  expected. Further, deviations from asymptotic Dirichlet boundary conditions are known to be described by irrelevant surface scaling fields $\lambda_j\propto 1/\cb_j$ (``extrapolation lengths'') that scale naively \cite{DDE83,Die86a}, so that their correction-to-scaling exponent is $\omega_\lambda=1$. Noting the degeneracy $\omega=\omega_\lambda=1$ at $d=3$,  standard RG considerations can be used to show that leading corrections to scaling $\propto L^{-1}\ln L$ along with those $\propto L^{-1}$ should occur for 
the effective amplitude $\tilde{\Delta}_{\mathrm{C}}(g,L)\equiv L^2[\fex(0,g,L)-2\fs(0,g)]$. As can be seen from the results for $\tilde{\Delta}_\mathrm{C}$ depicted in Fig.~\ref{fig:deltag}, convergence is very poor when $g$ is small, e.g., for $g=1$.
Relying on the results for  $\tilde{\Delta}_\mathrm{C}(1,L\lesssim 64)$, one could easily infer an incorrect value of $-0.026(1)$ for $\Delta_{\mathrm{C}}$, more than twice as large as the correct asymptotic one $\tilde\Delta_{\mathrm{C}}(1,\infty)=-0.0108(1)$, see Fig.~\ref{fig:deltag}. Convergence is far better for $g\gg 1$, as the minimum in $\tilde{\Delta}_\mathrm{C}$ is at $L_\mathrm{min} \etwa 80/g$. Fast convergence and high precision can be achieved by solving Eqs.~\eqref{eq:sctfex} with $\tau=0$ 
and $g$ set  to its fixed-point value\footnote{Recall that the bulk analog of the integral in Eq.~\eqref{eq:V-tau} contains a regularization-dependent term, which reads $-a(d)\rb$ in the notation of Ref.~\cite{MZ03}. Its coefficient $a(d)$ may be  identified as $1/g^*$ if $a(d)>0$. For our regularization method A, $a(d)=0$ as $n\to\infty$, so that $g^*=\infty$. For method B, $a(d)<0$. Hence no infrared-stable  $g^*>0$ exists and corrections to scaling cannot be suppressed in this manner; cf.\ Ref.~\cite[Sect.~2.4]{MZ03}.}  $g^*=\infty$, where the corrections $\propto L^{-1} \ln L$ vanish.

To extend this $g\to\infty$ analysis to $\tau\ne 0$, we note that the critical exponent $\nu=1$ and absorb the amplitude $\xi_+(g)=g/24\pi$ of the  bulk correlation length $\xi^{(+)}_{\mathrm{b}}=r_{\mathrm{b}}^{-1/2}\aseq \xi_+\tau^{-\nu}$ for $T>T_\mathrm{c}$ in the temperature variable, defining $t=\tau/\xi_+(g)$ so that the scaling variable becomes $x=t L$.
In this limit, Eqs.~\eqref{eq:sctfex} reduce to
\begin{subequations}\label{eq:sctfexinf}
\begin{equation}\label{eq:sct}
t=\langle z| \ln\mat{H} |z\rangle,
\end{equation}
\begin{equation}\label{eq:fbinf}
\fb(t) = \frac{1}{4\pi}
\begin{cases}
\sinh t - t & \text{for } t \geq 0,\\
0 & \text{for } t < 0,
\end{cases}
\end{equation}
\begin{equation}\label{eq:fexinf}
\fex(t,L)=\frac{1}{8\pi}\Tr\!\left[\mat{H} \left(1+t-\ln \mat{H} \right)\right]  - \frac{t L}{4\pi} - L \fb(t).
\end{equation}
\end{subequations}
From Eqs.~\eqref{eq:sctfexinf} we calculate the Casimir force $\beta \mathcal{F}_\mathrm{C}(t,L) = -\partial \fex(t,L)/\partial L\etwa -[\fex(t,L+1)-\fex(t,L-1)]/2$. 
To achieve the excellent data collapse shown in Fig.~\ref{fig:theta},
it turns out to be sufficient to write $\vartheta(x)\aseq \Leff^3 \, \beta \mathcal{F}_\mathrm{C}(t,L)$, introducing
an effective thickness $\Leff=L+\delta L$, as proposed in \cite{HasenbuschDeltaL} and substantiated by field theory \cite{DDE83}.

Inspection of Eq.~\eqref{eq:sct} reveals that the scaled lowest eigenvalue $\epsilon_0L^2$ is positive for all $x$ and vanishes $\asprop |x| \rme^{-|x|}$ as $x\to-\infty$. That is, for finite $L<\infty$, the system remains paramagnetic whenever $T>0$, due to the nonperturbative generation of a mass. Furthermore, the remaining eigenvalues $\epsilon_{\nu>0}$ approach the Neumann values. 

The numerical calculations are performed with 33 digits precision. This yields about $30$ significant digits in $\fex$.
For the effective thickness the form $\Leff= L+ \delta L + \sum_{i=1}^{m} b_i L^{-i}$ is chosen. The estimates of $\Delta_\mathrm{C}$ and $\delta L$ are then determined by analyzing $\fex(0,L)$ 
for $L=1600,1800,\ldots,3800$ and $L=4096$ with the ansatz 
$ \fex(0,L) = 2 f_s(0) + \Delta_\mathrm{C} \Leff^{-2}$.
Our final results, Eq.~\eqref{eq:Deltaest} and $\delta L=0.7255032704723(3)$, are obtained using $m=5$. As benchmark for the errors, the variations of the estimates resulting from analogous analyses with $m=4$ and different choices of thicknesses $L$ are used.

{\it Model B ---}
A simple cubic lattice model of sites $\bm{x}\equiv(\bm{y},z)\in\mathbb{Z}^d$ with $x_i=1,\dotsc,N_i$ is considered whose Hamiltonian follows from Eq.~\eqref{eq:Ham} through the replacements $\int\rmd^{d-1}y\int_0^L\rmd{z}\to \sum_{\bm{x}}$ and $(\nabla\bm{\phi})^2 \to  \sum_{i=1}^d|\bm{\phi}(\bm{x})-\bm{\phi}(\bm{x}+\bm{e}_i)|^2$, where  $\bm{e}_i$ are unit vectors along the principal axes. Along the $z$-direction, Dirichlet boundary conditions  are  again imposed; along all $y_i$-directions, periodic boundary conditions $\bm{\phi}_{\bm{y}}\equiv\bm{\phi}_{\bm{y}+N_i\bm{e}_i}$ are chosen and the limits $N_{i<d}\to\infty$ taken at fixed  $L\equiv N_d$. Just as in model A, the operator $\partial_z^2$ in Eqs.~\eqref{eq:f} and
\eqref{eq:ev} is replaced by its discrete analog $\mat{D}^2$. Owing to the discreteness of the lattice, the momentum integrations are restricted to $|p_i|\le \pi$. Further, to account for the modified dispersion relation, $4\sum_{i=1}^{d-1}\sin^2(p_i/2)$ must be substituted for $\bm{p}^2$. Thus the $\bm{p}$-integral of each series coefficient of the spectral sum $\sum_\nu$ in Eq.~\eqref{eq:V-tau} becomes an analytically computable $\epsilon_\nu$-dependent two-dimensional Watson integral. Its three-dimensional bulk analog can also be determined analytically \cite{JZ01}, along with their antiderivatives one encounters in the analogous momentum integrals of the right-hand side of Eq.~\eqref{eq:f}. 

For model B no fixed-point value $g^*>0$ exists to which $g$ could be set 
to eliminate leading scaling corrections \cite{MZ03}. Thus, scaling corrections 
$\propto L^{-1} \ln L$  remain for any value of $g$. However, the amplitudes of these 
corrections become minimal in the limit $g \rightarrow \infty$.  
We use $\Leff = L + a_0 \ln L+ \delta L + \sum_{i=1}^{m} (a_i \ln L + b_i) L^{-i} $ with $m=3$ to analyse $\fex(0,L)$, which we computed for various thicknesses
up to $L=4096$. We reproduce the value \eqref{eq:Deltaest} for the Casimir amplitude 
to 13 significant digits --- a striking confirmation of universality. Furthermore,
we get $a_0=-0.123903101(1)$ and $\delta L =0.81422072(1)$.  Analyzing our data also at  
the minimum of the thermodynamic Casimir force, we obtain results for $x_\mathrm{min}$ and 
$\vartheta_\mathrm{min}$ that are fully consistent with those obtained from model A.
The value of $a_0$  is consistent with the one obtained above from the analysis of $\fex(0,L)$. 
However, we get $\delta L=1.1979(1)$ and 
$\delta L=0.8924(1)$ from the location and the value of the minimum 
of the thermodynamic Casimir force, respectively. The plot of  $\vartheta$ in  
Fig.~\ref{fig:theta} uses $\Leff = L -0.1239031 \ln L + 1$, which results
in a good data collapse. However, we should keep in mind that this way corrections 
$\propto L^{-1}$ cannot be eliminated completely for the whole range of the scaling
argument $x$. 

Finally, let us comment on the stability of our results with respect to a change of  boundary conditions, as  investigated recently in \cite{DS11}. For both models A and B, the nearest-neighbor (NN) bonds $\beta J_j$  in the two boundary layers $j=1$ and $j=L$ were chosen to agree with the NN bonds  $\beta J$ in all other layers. Allowing for boundary couplings\footnote{A sc lattice model of type B with boundary bonds $J_1\ne J$ and $J_L\ne J$ can be mapped approximately onto the continuum field theory~\eqref{eq:Ham} with $c_1=1-2(d-1)(J_1/J-1)$ and $c_2=1-2(d-1)(J_L/J-1)$ \cite[p.~92]{Die86a}.} $J_1$ and $J_L\ne J$, one can change the corresponding diagonal elements of 
$\mat D^2$. However, this change is compensated by a corresponding change of the potential $\mat V$. As expected for $d=3$, the selfconsistent solutions remain asymptotically the same.

In summary, considering the $O(n)$ vector model on a slab with free boundary conditions, we expressed its universal scaling function for the Casimir force in the limit $n\to\infty$ exactly in terms of the eigensystem of the resulting selfconsistent 1d Schr\"odinger equation, which we then solved by numerical means using two qualitatively distinct regularization methods. We obtained consistent results that agree to many digits. They exhibit all qualitative features (i)--(iv) one expects to hold for general $n\ge2$, in particular, nontrivial crossovers from three-dimensional critical to two-dimensional pseudo-critical behavior and to the low-temperature behavior dictated by confined Goldstone modes.\footnote{In the $O(2)$ case, a Kosterlitz-Thouless transition occurs at a temperature $T_\mathrm{KT}$ (\cite{KT2},  and refs. therein),
which is the inflection point below the location of the minimum of the Casimir force, cf. \cite{KT}. This is absent when $n>2$ and hence for $n\to\infty$. For $n=2$, it is barely detectable in $\vartheta$; see Refs. \cite{MC1,MC3}} Besides being interesting in their own right, these results could pave the way to successful approximate analytical treatments of such challenging problems. We expect them to play a role as fruitful as large-$n$ solutions have done in the theory of quantum critical phenomena \cite{Sac11}. Furthermore, our analysis can be extended in a straightforward fashion to include bulk and surface magnetic fields and to study appropriate quantum versions of the model.

Some of us (DG, AH, FMS) like to thank Denis Comtesse for fruitful discussions.
We gratefully acknowledge partial support by DFG for two of us (HWD and FMS) via grant DI 378/5 and for one of us (MH) via grant HA 3150/2-2. SBR was partially supported by the Belarusian Foundation for Fundamental Research.


\begin{thebibliography}{19}
\expandafter\ifx\csname natexlab\endcsname\relax\def\natexlab#1{#1}\fi
\expandafter\ifx\csname bibnamefont\endcsname\relax
  \def\bibnamefont#1{#1}\fi
\expandafter\ifx\csname bibfnamefont\endcsname\relax
  \def\bibfnamefont#1{#1}\fi
\expandafter\ifx\csname citenamefont\endcsname\relax
  \def\citenamefont#1{#1}\fi
\expandafter\ifx\csname url\endcsname\relax
  \def\url#1{\texttt{#1}}\fi
\expandafter\ifx\csname urlprefix\endcsname\relax\def\urlprefix{URL }\fi
\providecommand{\bibinfo}[2]{#2}
\providecommand{\eprint}[2][]{\url{#2}}

\bibitem{Cas48}
\bibinfo{author}{\bibfnamefont{H.~B.~G.} \bibnamefont{Casimir}},
\bibinfo{journal}{Proc. K. Ned. Akad. Wet.} \textbf{\bibinfo{volume}{B51}},
\bibinfo{pages}{793} (\bibinfo{year}{1948}).

\bibitem{casiqmrev}
\bibinfo{author}{\bibfnamefont{M.}~\bibnamefont{Bordag}},
  \bibinfo{author}{\bibfnamefont{U.}~\bibnamefont{Mohideen}}, \bibnamefont{and}
  \bibinfo{author}{\bibfnamefont{V.~M.} \bibnamefont{Mostepanenko}},
  \bibinfo{journal}{Phys. Rep.} \textbf{\bibinfo{volume}{353}},
  \bibinfo{pages}{1} (\bibinfo{year}{2001}).

\bibitem{FdG78}
\bibinfo{author}{\bibfnamefont{M.~E.} \bibnamefont{Fisher}} \bibnamefont{and}
  \bibinfo{author}{\bibfnamefont{P.-G.} \bibnamefont{de~Gennes}},
  \bibinfo{journal}{C.\ R.\ S{\'e}ances.\ Acad.\ Sci.\ S{\'e}rie B}
  \textbf{\bibinfo{volume}{287}}, \bibinfo{pages}{207} (\bibinfo{year}{1978}).

\bibitem{casitdrev}

\bibinfo{author}{\bibfnamefont{M.}~\bibnamefont{Krech}},
  \emph{\bibinfo{title}{{C}asimir Effect in Critical Systems}}
  (\bibinfo{publisher}{World Scientific}, \bibinfo{address}{Singapore},
  \bibinfo{year}{1994}) and
\bibinfo{author}{\bibfnamefont{J.~G.} \bibnamefont{Brankov}},
  \bibinfo{author}{\bibfnamefont{D.~M.} \bibnamefont{Dantchev}},
  \bibnamefont{and} \bibinfo{author}{\bibfnamefont{N.~S.}
  \bibnamefont{Tonchev}}, \emph{\bibinfo{title}{Theory of Critical Phenomena in
  Finite-Size Systems --- Scaling and Quantum Effects}}
  (\bibinfo{publisher}{World Scientific}, \bibinfo{address}{Singapore},
  \bibinfo{year}{2000}); 
  \bibinfo{author}{\bibfnamefont{A.}~\bibnamefont{Gambassi}},
  \bibinfo{journal}{J. Phys.: Conference Series}
  \textbf{\bibinfo{volume}{161}}, \bibinfo{pages}{012037}
  (\bibinfo{year}{2009}).

\bibitem{tdc} 
\bibinfo{author}{\bibfnamefont{G.~L.} \bibnamefont{Klimchitskaya}},
  \bibinfo{author}{\bibfnamefont{U.}~\bibnamefont{Mohideen}}, \bibnamefont{and}
  \bibinfo{author}{\bibfnamefont{V.~M.} \bibnamefont{Mostepanenko}},
  \bibinfo{journal}{Rev. Mod. Phys.} \textbf{\bibinfo{volume}{81}},
  \bibinfo{pages}{1827} (\bibinfo{year}{2009}).

\bibitem{KG99}
\bibinfo{author}{\bibfnamefont{M.}~\bibnamefont{Kardar}} \bibnamefont{and}
  \bibinfo{author}{\bibfnamefont{R.}~\bibnamefont{Golestanian}},
  \bibinfo{journal}{Rev. Mod. Phys.} \textbf{\bibinfo{volume}{71}},
  \bibinfo{pages}{1233} (\bibinfo{year}{1999}).

\bibitem{GC99}
\bibinfo{author}{\bibfnamefont{R.}~\bibnamefont{Garcia}} \bibnamefont{and}
  \bibinfo{author}{\bibfnamefont{M.~H.~W.} \bibnamefont{Chan}},
  \bibinfo{journal}{Phys. Rev. Lett.} \textbf{\bibinfo{volume}{83}},
  \bibinfo{pages}{1187} (\bibinfo{year}{1999});
\bibinfo{author}{\bibfnamefont{A.}~\bibnamefont{Ganshin}},
\bibinfo{author}{\bibfnamefont{S.}~\bibnamefont{Scheidemantel}},
\bibinfo{author}{\bibfnamefont{R.}~\bibnamefont{Garcia}},
\bibnamefont{and}
  \bibinfo{author}{\bibfnamefont{M.~H.~W.} \bibnamefont{Chan}},
  \bibinfo{journal}{Phys. Rev. Lett.} \textbf{\bibinfo{volume}{97}},
  \bibinfo{pages}{075301} (\bibinfo{year}{2006}).

\bibitem{HHGDB08}
\bibinfo{author}{\bibfnamefont{C.}~\bibnamefont{Hertlein}},
  \bibinfo{author}{\bibfnamefont{L.}~\bibnamefont{Helden}},
  \bibinfo{author}{\bibfnamefont{A.}~\bibnamefont{Gambassi}},
  \bibinfo{author}{\bibfnamefont{S.}~\bibnamefont{Dietrich}}, \bibnamefont{and}
  \bibinfo{author}{\bibfnamefont{C.}~\bibnamefont{Bechinger}},
  \bibinfo{journal}{Nature} \textbf{\bibinfo{volume}{451}},
  \bibinfo{pages}{172} (\bibinfo{year}{2008}).

\bibitem{MC1}
\bibinfo{author}{\bibfnamefont{A.}~\bibnamefont{Hucht}},
  \bibinfo{journal}{Phys. Rev. Lett.} \textbf{\bibinfo{volume}{99}},
  \bibinfo{eid}{185301} (pages~\bibinfo{numpages}{4}) (\bibinfo{year}{2007}).

\bibitem{MC2}
\bibinfo{author}{\bibfnamefont{O.}~\bibnamefont{Vasilyev}},
  \bibinfo{author}{\bibfnamefont{A.}~\bibnamefont{Gambassi}},
  \bibinfo{author}{\bibfnamefont{A.}~\bibnamefont{Macio{\l}ek}},
  \bibnamefont{and} \bibinfo{author}{\bibfnamefont{S.}~\bibnamefont{Dietrich}},
  \bibinfo{journal}{Europhys. Lett.} \textbf{\bibinfo{volume}{80}},
  \bibinfo{pages}{60009 (6pp)} (\bibinfo{year}{2007}).

\bibitem{MC3}
\bibinfo{author}{\bibfnamefont{M.}~\bibnamefont{Hasenbusch}},
  \bibinfo{journal}{Journal of Statistical Mechanics: Theory and Experiment}
  \textbf{\bibinfo{volume}{2009}}, \bibinfo{pages}{P07031}
  (\bibinfo{year}{2009});
\bibinfo{author}{\bibfnamefont{M.}~\bibnamefont{Hasenbusch}},
\bibinfo{journal}{Phys. Rev. B}
\textbf{\bibinfo{volume}{81}},\bibinfo{pages}{165412}
(\bibinfo{year}{2010}).

\bibitem{ZSRKC07}
\bibinfo{author}{\bibfnamefont{R.}~\bibnamefont{Zandi}},
  \bibinfo{author}{\bibfnamefont{A.}~\bibnamefont{Shackell}},
  \bibinfo{author}{\bibfnamefont{J.}~\bibnamefont{Rudnick}},
  \bibinfo{author}{\bibfnamefont{M.}~\bibnamefont{Kardar}}, \bibnamefont{and}
  \bibinfo{author}{\bibfnamefont{L.~P.} \bibnamefont{Chayes}},
  \bibinfo{journal}{Physical Review E (Statistical, Nonlinear, and Soft Matter
  Physics)} \textbf{\bibinfo{volume}{76}}, \bibinfo{eid}{030601}
  (pages~\bibinfo{numpages}{4}) (\bibinfo{year}{2007}).
  
\bibitem{MGD07}
\bibinfo{author}{\bibfnamefont{A.}~\bibnamefont{Macio{\l}ek}},
  \bibinfo{author}{\bibfnamefont{A.}~\bibnamefont{Gambassi}}, \bibnamefont{and}
  \bibinfo{author}{\bibfnamefont{S.}~\bibnamefont{Dietrich}},
  \bibinfo{journal}{Physical Review E (Statistical, Nonlinear, and Soft Matter
  Physics)} \textbf{\bibinfo{volume}{76}}, \bibinfo{eid}{031124}
  (pages~\bibinfo{numpages}{17}) (\bibinfo{year}{2007}).

  
\bibitem{Kno73}
\bibinfo{author}{\bibfnamefont{H.~J.~F.} \bibnamefont{Knops}},
  \bibinfo{journal}{J. Math. Phys.} \textbf{\bibinfo{volume}{14}},
  \bibinfo{pages}{1918} (\bibinfo{year}{1973}).


\bibitem{freeNbc}
\bibinfo{author}{\bibfnamefont{H.}~\bibnamefont{Li}} \bibnamefont{and}
  \bibinfo{author}{\bibfnamefont{M.}~\bibnamefont{Kardar}},
  \bibinfo{journal}{Phys. Rev. A} \textbf{\bibinfo{volume}{46}},
  \bibinfo{pages}{6490} (\bibinfo{year}{1992}).
  
\bibitem{lowTNbc}
\bibinfo{author}{\bibfnamefont{H.~W.} \bibnamefont{Diehl}} \bibnamefont{and}
  \bibinfo{author}{\bibfnamefont{A.}~\bibnamefont{N{\"u}sser}},
  \bibinfo{journal}{Phys. Rev. Lett.} \textbf{\bibinfo{volume}{56}},
  \bibinfo{pages}{2834} (\bibinfo{year}{1986}), and
\bibinfo{author}{\bibfnamefont{H.~W.} \bibnamefont{Diehl}},
  \bibinfo{journal}{Phys.\ Lett.} \textbf{\bibinfo{volume}{75A}},
  \bibinfo{pages}{375} (\bibinfo{year}{1980})

\bibitem{we}
\bibinfo{note}{Details will be published elsewhere.}

  \bibitem{BM77}
\bibinfo{author}{\bibfnamefont{A.~J.} \bibnamefont{Bray}} \bibnamefont{and}
  \bibinfo{author}{\bibfnamefont{M.~A.} \bibnamefont{Moore}},
  \bibinfo{journal}{J.\ Phys.\ A} \textbf{\bibinfo{volume}{10}},
  \bibinfo{pages}{1927} (\bibinfo{year}{1977}).

  \bibitem{Dan98}
\bibinfo{author}{\bibfnamefont{D.~M.} \bibnamefont{Danchev}},
  \bibinfo{journal}{Phys. Rev. E} \textbf{\bibinfo{volume}{58}},
  \bibinfo{pages}{1455} (\bibinfo{year}{1998}).
  
  \bibitem{Cometal}
\bibinfo{author}{\bibfnamefont{D.}~\bibnamefont{Comtesse}},
 (\bibinfo{year}{2008}),
  \bibinfo{note}{{D}iplomarbeit, Universit{\"a}t Duis\-burg-Essen (2008)} and  
\bibinfo{author}{\bibfnamefont{D.}~\bibnamefont{Comtesse}},
  \bibinfo{author}{\bibfnamefont{A.}~\bibnamefont{Hucht}}, \bibnamefont{and}
  \bibinfo{author}{\bibfnamefont{D.}~\bibnamefont{Gr{\"u}neberg}}, arXiv:0904.3661v2.


\bibitem{DDE83}
\bibinfo{author}{\bibfnamefont{H.~W.} \bibnamefont{Diehl}},
  \bibinfo{author}{\bibfnamefont{S.}~\bibnamefont{Dietrich}}, \bibnamefont{and}
  \bibinfo{author}{\bibfnamefont{E.}~\bibnamefont{Eisenriegler}},
  \bibinfo{journal}{Phys. Rev. B} \textbf{\bibinfo{volume}{27}},
  \bibinfo{pages}{2937} (\bibinfo{year}{1983}); 

\bibitem{Die86a}
\bibinfo{author}{\bibfnamefont{H.~W.} \bibnamefont{Diehl}}, in
  \emph{\bibinfo{booktitle}{Phase Transitions and Critical Phenomena}}, edited
  by \bibinfo{editor}{\bibfnamefont{C.}~\bibnamefont{Domb}} \bibnamefont{and}
  \bibinfo{editor}{\bibfnamefont{J.~L.} \bibnamefont{Lebowitz}}
  (\bibinfo{publisher}{Academic}, \bibinfo{address}{London},
  \bibinfo{year}{1986}), vol.~\bibinfo{volume}{10}, pp.
  \bibinfo{pages}{75--267}; \bibinfo{journal}{Int.\ J.\ Mod.\ Phys.\ B} \textbf{\bibinfo{volume}{11}},
  \bibinfo{pages}{3503} (\bibinfo{year}{1997}).
  
\bibitem{MZ03}
\bibinfo{author}{\bibfnamefont{M.}~\bibnamefont{Moshe}} \bibnamefont{and}
  \bibinfo{author}{\bibfnamefont{J.}~\bibnamefont{Zinn-Justin}},
  \bibinfo{journal}{Phys. Rep.} \textbf{\bibinfo{volume}{385}},
  \bibinfo{pages}{69} (\bibinfo{year}{2003}).
  
\bibitem{HasenbuschDeltaL}
  \bibinfo{note}{T. W. Capehart and M. E. Fisher, Phys. Rev. B \textbf{13}, 5021  (1976).}
    
\bibitem{JZ01}
  \bibinfo{author}{\bibfnamefont{G.~S.} \bibnamefont{Joyce}} \bibnamefont{and}
  \bibinfo{author}{\bibfnamefont{I.~J.} \bibnamefont{Zucker}},
  \bibinfo{journal}{J. Phys. A} \textbf{\bibinfo{volume}{34}},
  \bibinfo{pages}{7349} (\bibinfo{year}{2001}).
 
\bibitem{KT2}
\bibinfo{author}{\bibfnamefont{M.}~\bibnamefont{Hasenbusch}},
  \bibinfo{journal}{Journal of Statistical Mechanics: Theory and Experiment}
  \textbf{\bibinfo{volume}{2009}}, \bibinfo{pages}{P02005}
  (\bibinfo{year}{2009})

\bibitem{KT}
\bibinfo{author}{\bibfnamefont{A.}~\bibnamefont{Hucht}},
  \bibinfo{author}{\bibfnamefont{D.}~\bibnamefont{Gr\"uneberg}}, \bibnamefont{and}
  \bibinfo{author}{\bibfnamefont{F.~M.}~\bibnamefont{Schmidt}},
  \bibinfo{journal}{Phys. Rev. E} \textbf{\bibinfo{volume}{83}}, \bibinfo{eid}{051101} (\bibinfo{year}{2011}).

\bibitem{DS11}
\bibinfo{author}{\bibfnamefont{F.~M.} \bibnamefont{Schmidt}} \bibnamefont{and}
  \bibinfo{author}{\bibfnamefont{H.~W.} \bibnamefont{Diehl}},
  \bibinfo{journal}{Phys. Rev. Lett.} \textbf{\bibinfo{volume}{101}},
  \bibinfo{eid}{100601} (\bibinfo{year}{2008});
\bibinfo{author}{\bibfnamefont{H.~W.} \bibnamefont{Diehl}} \bibnamefont{and}
  \bibinfo{author}{\bibfnamefont{F.~M.} \bibnamefont{Schmidt}},
  \bibinfo{journal}{New J. Phys.} \textbf{\bibinfo{volume}{13}},
  \bibinfo{pages}{123025} (\bibinfo{year}{2011}).

\bibitem{Sac11}
\bibinfo{author}{\bibfnamefont{S.}~\bibnamefont{Sachdev}},
  \emph{\bibinfo{title}{Quantum Phase Transitions}}
  (\bibinfo{publisher}{Cambridge University Press}, \bibinfo{address}{Cambridge
  (UK)}, \bibinfo{year}{2011}), \bibinfo{edition}{2nd} ed.
  
 
\end{thebibliography}
\end{document}